\documentclass[12pt]{article}

\usepackage{enumitem}
\usepackage{float}
\usepackage{cite}
\usepackage{amsfonts}
\usepackage{amssymb}
\usepackage{amsmath}
\usepackage{xcolor}
\usepackage{mathtools}

\usepackage{graphicx}

\def\gtwid{\mathrel{\raise.3ex\hbox{$>$\kern-.75em\lower1ex\hbox{$\sim$}}}}
\def\ltwid{\mathrel{\raise.3ex\hbox{$<$\kern-.75em\lower1ex\hbox{$\sim$}}}}
\def\square{\kern1pt\vbox{\hrule height 1.2pt\hbox{\vrule width 1.2pt\hskip 3pt
   \vbox{\vskip 6pt}\hskip 3pt\vrule width 0.6pt}\hrule height 0.6pt}\kern1pt}

\begin{document}

\begin{titlepage}

\begin{flushright}
UFIFT-QG-24-08
\end{flushright}

\vskip 2cm

\begin{center}
{\bf Resumming Photon Loops for Inflationary Gravity}
\end{center}

\vskip 0.5cm

\begin{center}
A. J. Foraci$^{*}$ and R. P. Woodard$^{\dagger}$
\end{center}

\vskip 0.5cm

\begin{center}
\it{Department of Physics, University of Florida,\\
Gainesville, FL 32611, UNITED STATES}
\end{center}

\vspace{0.5cm}

\begin{center}
ABSTRACT
\end{center}
A previous calculation of the 1-loop photon contribution to the graviton 
self-energy on de Sitter background is considered. We first show that 
there is no local obstacle to conservation, unlike the contribution from
a loop of massless, minimally coupled scalars. This is correlated to the
absence of an Eddington ($R^2$) counterterm and to the vanishing of the
stress tensor when the photon in integrated out in the presence of a 
constant graviton field. We also show that there is a secularly growing
1-loop contribution to the electric components of the Weyl tensor for
plane wave gravitons. Its coefficient agrees with that of the secular 
1-loop correction to the Newtonian potential, and both can be resummed
using a variant of the renormalization group. 

\begin{flushleft}
PACS numbers: 04.50.Kd, 95.35.+d, 98.62.-g
\end{flushleft}

\vskip 2cm

\begin{flushleft}
$^{*}$ e-mail: aforaci@ufl.edu \\
$^{\dagger}$ e-mail: woodard@phys.ufl.edu
\end{flushleft}

\end{titlepage}

\section{Introduction}

The background geometry of cosmology can be expressed either in
terms of co-moving time $t$ or conformal time $\eta$,
\begin{equation}
ds^2 = -dt^2 + a^2(t) d\vec{x} \!\cdot\! d\vec{x} = a^2 [-d\eta^2
+ d\vec{x} \!\cdot\! d\vec{x}] \; . \label{background}
\end{equation}
Derivatives of the scale factor $a(t)$ give the Hubble parameter $H(t)$ 
and the first slow roll parameter $\epsilon(t)$,
\begin{equation}
H(t) \equiv \frac{\dot{a}}{a} \qquad , \qquad \epsilon(t) \equiv
-\frac{\dot{H}}{H^2} \; . \label{parameters}
\end{equation}
When $\dot{a}$ and $\ddot{a}$ are both positive (which means $H > 0$
and $0 \leq \epsilon < 1$) the universe is said to be inflating. The
epoch of primordial inflation is conjectured to have occurred during
the earliest instants of cosmic history, with a Hubble parameter as
high as $10^{14}$ GeV, and $\epsilon < 0.003$ \cite{Tristram:2020wbi}.
Owing to the very small first slow roll parameter it is for many 
purposes reasonable to approximate the geometry of primordial inflation
as de Sitter, with constant $H$, vanishing $\epsilon$ and scale factor
$a(t) = e^{H t} = -\frac1{H \eta}$.

The accelerated expansion of primordial inflation rips virtual, long
wavelength gravitons out of the vacuum \cite{Starobinsky:1979ty}. 
Because more and more of these quanta are produced as inflation progresses,
graviton loop corrections which would be constant on flat space background
can become time dependent. For example, when the 1-loop graviton contribution
to the vacuum polarization on de Sitter background \cite{Leonard:2013xsa} is 
used to quantum-correct Maxwell's equation, the response to a static point 
charge becomes \cite{Glavan:2013jca},
\begin{equation}
\Phi(t,r) = \tfrac{Q}{4\pi \varepsilon_0 a(t) r} \Bigl\{1 + \tfrac{2 \hbar G}{
3 \pi c^3 a^2(t) r^2} + \tfrac{2 \hbar G H^2}{\pi c^5} \ln[\tfrac{a(t) H r}{c}]
+ \dots \Bigr\} \; . \label{Coulomb}
\end{equation}
The fractional correction proportional to $G/r^2$ is the de Sitter version
of an effect first seen on flat space background in 1970 \cite{Radkowski:1970}.
The secular effects proportional to $G H^2$ derive from cosmological particle
production.  

During a prolonged period of inflation these secular corrections can become
large enough that perturbation theory breaks down. In order to understand 
what happens at later time one must employ some kind of nonperturbative
resummation technique. The struggle to develop such a technique has been 
long and confusing because secular corrections have two distinct sources,
each requiring its own resummation technique. The first source is from what 
DeWitt and Brehme termed the ``tail'' part of the graviton propagator 
\cite{DeWitt:1960fc}. Because the mode functions of dynamical gravitons agree
with those of the massless, minimally coupled scalar \cite{Lifshitz:1945du}
their propagators agree as well. The tail is the logarithm term of this
propagator on $D=4$ dimensional de Sitter background,
\begin{equation}
i\Delta(x;x') = \tfrac1{4\pi^2}[ \tfrac1{a a' \Delta x^2} - \tfrac{H^2}{8\pi^2}
\ln[\tfrac14 H^2 \Delta x^2] \; , \label{tailterm}
\end{equation}
where $\Delta x^2 \equiv (x - x')^{\mu} (x - x')^{\nu} \eta_{\mu\nu}$ is 
the conformal coordinate interval. Secular corrections from this source can be 
resummed using a variant of Starobinsky's stochastic formalism 
\cite{Starobinsky:1986fx,Starobinsky:1994bd,Tsamis:2005hd,Miao:2021gic}.

The other source of secular logarithms comes from renormalization. To make 
the discussion  concrete, suppose we employ dimensional regularization in
conformal coordinates, in which case interaction vertices inherit a factor
of $a^D$ from the measure $\sqrt{-g}$. It turns out that the propagator from
$x^{\mu}$ to ${x'}^{\mu}$ goes like $1/(a a')^{D/2 -1}$, times integer powers
of $a a'$, so that $D$-dependent scale factors cancel from 1-loop diagrams,
in both the ultraviolet divergent and finite parts. On the other hand,
counterterms inherit $D$-dependent powers of $a$ from the measure factor, 
so there is an incomplete cancellation between primitive divergences and
counterterms \cite{Miao:2021gic},
\begin{equation}
\tfrac{(2 H)^{D-4}}{D - 4} - \tfrac{(a \mu)^{D-4}}{D - 4} = 
-\ln(\tfrac{\mu a}{2 H}) + O(D \!-\! 4) \; . \label{RGsource}
\end{equation}
Here $\mu$ is the usual mass scale of dimensional regularization. The close
relation between $\mu$ and $a(t)$ in relation (\ref{RGsource}) suggests that
this second source of secular logarithms can be resummed by a variant of the
renormalization group \cite{Miao:2021gic}.

Theories with derivative interactions generally receive large loop
corrections from both sources. One example is gravity plus a massless, 
minimally coupled scalar. When a loop of gravitons is included in the scalar
self-mass, and then used to quantum-correct the linearized effective field
equation, the late time response to a stationary point source $J(t,\vec{x}) =
K a(t) \delta^3(\vec{x})$ approaches the time-independent constant 
\cite{Glavan:2021adm},
\begin{equation}
\Phi(t,r) \longrightarrow \tfrac{HK}{4\pi} \ln(\tfrac{Hr}{c}) \Bigl\{1 -
\tfrac{2 \hbar G H^2}{\pi c^5} \ln(\tfrac{Hr}{c}) + \dots \Bigr\} \; .
\label{scalarpot}
\end{equation}
That logarithm turns out to derive from the second source (\ref{RGsource})
\cite{Glavan:2021adm}. However, one can also include the scalar loop 
contribution to the graviton self-energy, and then use that to quantum-correct
the linearized Einstein equation. This induces changes in both the electric
components of the Weyl tensor for gravitational radiation and in the response
to a static point mass \cite{Miao:2024atw},
\begin{eqnarray}
C_{0i0j}(t,\vec{x}) &\!\!\! = \!\!\!& C^{\rm tree}_{0i0j}(t,\vec{x}) \Bigl\{1 
- \tfrac{3 \hbar G H^2}{10 \pi c^5} \ln[a(t)] + \dots \Bigr\} \; , \qquad
\label{MMCrad} \\
\Psi(t,r) &\!\!\! = \!\!\!& \tfrac{G M}{a(t) r} \Bigl\{1 + \tfrac{\hbar G}{20 
\pi c^3 a^2(t) r^2} - \tfrac{3 \hbar G H^2}{10 \pi c^5} \ln[\tfrac{a(t) H r}{c}]
+ \dots \Bigr\} \; . \qquad \label{MMCNewton}
\end{eqnarray}
These logarithms also derive from the second source (\ref{RGsource}) and can
be resummed using a variant of the renormalization group to give 
\cite{Miao:2024nsz},
\begin{eqnarray}
C_{0i0j}(t,\vec{x}) &\!\!\! \longrightarrow \!\!\!& C^{\rm tree}_{0i0j}(t,\vec{x})
\!\times\! [a(t)]^{-\frac{3 \hbar G H^2}{10 \pi c^5}} \; , \qquad
\label{MMCresumrad} \\
\Psi(t,r) &\!\!\! \longrightarrow \!\!\!& \tfrac{G M}{a(t) r} \!\times\!
[\tfrac{a(t) H r}{c}]^{-\frac{3 \hbar G H^2}{10 \pi c^5}} \; . \qquad 
\label{MMCresumNewton}
\end{eqnarray}
However, this model does show an important ``tail'' effect in the form of an 
induced 1-point function which must be removed by a finite renormalization
of the cosmological constant in order to make the graviton self-energy
conserved \cite{Tsamis:2023fri}.

The current work aims to extend this analysis to electromagnetic contributions
to the graviton self-energy. A result for this was obtained previously 
\cite{Wang:2015eaa}, but conservation was not checked carefully for the sort of 
delta function obstacle which the scalar model manifests \cite{Tsamis:2023fri}. 
The potentially nonconserved result was instead expressed using manifestly 
conserved structure functions and used to solve for quantum corrections to the 
Newtonian potential. When this same procedure was employed for the scalar model 
\cite{Leonard:2014zua} (whose conservation obstacle had not then been recognized), 
the resulting 1-loop correction to the Newtonian potential disagrees slightly with 
the correct result (\ref{MMCNewton}) \cite{Park:2015kua}. One of our goals is to
determine whether or not electromagnetic contributions to the graviton self-energy
manifest a similar obstacle to conservation. We also extend the solution of the
graviton mode function to compute secular corrections to the electric components
of the Weyl tensor of gravitational radiation. Finally, we employ the 
renormalization group to resum the large loop corrections.

This paper contains six sections, of which 2-5 employ the standard 
$\hbar = 1 = c$ units of particle theory. In section 2 we review primitive 
contributions to the graviton self-energy with the object of obtaining a form that 
can be used to check for delta function obstacles to conservation. In section 3 
that check is performed, with the result that no such obstacle exists. We also 
show that this is correlated with the absence of a tail-induced effective stress
tensor. Section 4 solves the linearized effective field equation for gravitational 
radiation to obtain a result analogous to (\ref{MMCrad}). Section 5 shows that this 
result, and the previous result for the Newtonian potential \cite{Wang:2015eaa}, 
which we now know to be correct, can be resumed using a variant of the 
renormalization group. Our conclusions comprise section 6.

\section{Previous Result for $-i [\mbox{}^{\mu\nu} \Sigma^{\rho\sigma}](x;x')$}

The purpose of this section is to review the primitive result for the 1-loop
electromagnetic contributions to the graviton self-energy, so that we can 
demonstrate conservation in the next section. We begin by specifying the 
diagrams and giving the vertices. We then reduce the two primitive contributions
in terms of the gauge invariant field strength correlator.

\begin{figure}[H]
\centering
\vskip 1cm
\includegraphics[width=8cm]{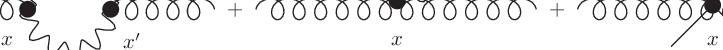}
\caption{\footnotesize Electromagnetic diagrams which contribute to the 1-loop
graviton self-energy. Wavy lines stand for photons and curly lines for gravitons.}
\label{diagrams}
\end{figure}

The graviton self-energy $-i [\mbox{}^{\mu\nu} \Sigma^{\rho\sigma}](x;x')$ is the
1PI (which stands for one-particle-irreducible) 2-graviton function. The primitive 
1-loop contributions to it from electromagnetism are the first two diagrams of 
Figure~\ref{diagrams}. They follow from the coupling of electromagnetism to 
gravity which is given by the Lagrangian,
\begin{equation}
\mathcal{L}_{\text{EM}} = -\tfrac{1}{4} F_{\mu\nu} F_{\rho\sigma} g^{\mu\rho}
g^{\nu\sigma}\sqrt{-g} \; . \label{EMaction}
\end{equation}
The action is the integral of this Lagrangian. We define the graviton field by 
conformally rescaling the metric in conformal coordinates,
\begin{equation}
g_{\mu\nu}(x) \equiv a^2 [\eta_{\mu\nu} + \kappa h_{\mu\nu}] \qquad , \qquad
\kappa^2 \equiv 16 \pi G \; . \label{graviton}
\end{equation}
Performing functional differentiation of the action once (twice) with respect 
to the graviton field, and then setting $h_{\mu\nu} = 0$, yields the 3-point 
(4-point) vertex interactions. Then the 3-point (4-point) contributions to the 
1PI 2-point function are constructed from two (one) of these vertices 
\cite{Wang:2015eaa},
\begin{align}
    -i\bigl[{}^{\mu\nu}\Sigma^{\rho\sigma}_{\text{3pt}}\bigr](x;x')
& = \frac12
\left(-i\kappa\right)a^{D-4}V^{\alpha\kappa\gamma\theta\mu\nu} \partial_{\kappa}
\partial'_{\lambda} i \bigl[{}_{\alpha} \Delta_{\beta}\bigr](x;x') \nonumber \\
& \times\left( -i \kappa \right) a'^{D-4} V^{\beta\lambda\delta\phi\rho\sigma}
\partial_{\theta} \partial'_{\phi} i \bigl[{}_{\gamma} \Delta_{\delta}\bigr](x;x') 
\; , \label{3ptExpr} \\[10pt]
-i \bigl[{}^{\mu\nu} \Sigma^{\rho\sigma}_{\text{4pt}}\bigr](x;x')
& = \left(-i\kappa^2\right) a^{D-4} U^{\alpha\kappa\beta\lambda\mu\nu\rho\sigma}
\partial_{\kappa} \partial'_{\lambda} i \bigl[{}_{\alpha} \Delta_{\beta}\bigr](x;x')
\delta^D(x-x') \; . \label{4ptExpr}
\end{align}
Here, the two constant tensor factors from these vertices are,
\begin{eqnarray}
\lefteqn{V^{\alpha\kappa\beta\lambda\mu\nu}
=\eta^{\mu\nu}\eta^{\alpha[\beta}\eta^{\lambda]\kappa}
+4\eta^{\mu)[\alpha}\eta^{\kappa][\beta}\eta^{\lambda](\nu} \; ,} \label{Vvert} \\
\lefteqn{U^{\alpha\kappa\beta\lambda\mu\nu\rho\sigma}
= (\tfrac14 \eta^{\mu\nu} \eta^{\rho\sigma} \!-\! \tfrac12 \eta^{\mu(\rho} 
\eta^{\sigma)\nu} ) \eta^{\alpha[\beta} \eta^{\lambda]\kappa} + \eta^{\mu\nu} 
\eta^{\rho)[\alpha} \eta^{\kappa][\beta} \eta^{\lambda](\sigma} } \nonumber \\
& & \hspace{-0.5cm} + \eta^{\rho\sigma} \eta^{\mu)[\alpha} \eta^{\kappa][\beta} 
\eta^{\lambda](\nu} \!+\! \eta^{\alpha(\mu} \eta^{\nu)(\rho} \eta^{\sigma)[\beta}
\eta^{\lambda]\kappa} \!+\! \eta^{\kappa(\mu} \eta^{\nu)(\rho} \eta^{\sigma)[\lambda}
\eta^{\beta]\alpha} \!+\! \eta^{\alpha(\mu} \eta^{\nu)[\beta} \eta^{\lambda](\rho}
\eta^{\sigma)\kappa} \nonumber \\
& & \hspace{1cm} + \eta^{\kappa(\mu} \eta^{\nu)[\lambda} \eta^{\beta](\rho}
\eta^{\sigma)\alpha} + \eta^{\alpha[\beta} \eta^{\lambda](\mu} \eta^{\nu)(\rho}
\eta^{\sigma)\kappa} + \eta^{\kappa[\lambda} \eta^{\beta](\mu} \eta^{\nu)(\rho}
\eta^{\sigma)\alpha} \; . \qquad \label{Uvert}
\end{eqnarray}
Parenthesized indices are symmetrized and indices enclosed in square brackets are
anti-symmetrized. 

The numerous anti-symmetrizations in expressions (\ref{Vvert}-\ref{Uvert}) suggest
that the 3-point and 4-point contributions are more conveniently expressed in terms 
of a four-index field-strength correlator, 
\begin{equation}
i \bigl[{}_{\alpha\beta} \Delta_{\gamma\delta}\bigr](x;x') \equiv
\langle \Omega \vert F_{\alpha\beta}(x) F_{\gamma\delta}(x') \vert \Omega \rangle 
= 4 \partial_{[\alpha}\partial'_{[[\gamma} i \bigl[{}_{\beta]} 
\Delta_{\delta]]}\bigr](x;x') \; . \label{FFcor}
\end{equation}
Because the field strength is $U(1)$ gauge invariant, there is no dependence on the
electromagnetic gauge fixing. In this notation, the 3-point and 4-point contributions 
to the graviton self-energy become,
\begin{eqnarray}
\lefteqn{
-i \bigl[{}^{\mu\nu} \Sigma^{\rho\sigma}_{\text{4pt}} \bigr](x;x')
= -\tfrac{\kappa^2}{2} a'^{D-4} i\delta^D(x-x') \Bigl\{ (\tfrac18 \eta^{\mu\nu}
\eta^{\rho\sigma} \!-\! \tfrac14 \eta^{\mu(\rho} \eta^{\sigma)\nu} ) } 
\nonumber \\
& & \hspace{1cm} \times i \bigl[{}^{\alpha\beta} \Delta_{\alpha\beta}\bigr](x';x') 
- \tfrac12 \eta^{\mu\nu} i\bigl[{}^{\rho\beta} 
\Delta^{\sigma}{}_{\beta}\bigr](x';x') \!-\! \tfrac12 \eta^{\rho\sigma} 
i\bigl[{}^{\mu\beta} \Delta^{\nu}{}_{\beta}\bigr](x';x') \nonumber\\
& & \hspace{3cm} + 2 \eta^{\nu) (\rho} i\bigl[{}^{\sigma)\beta} 
\Delta^{(\mu}{}_{\beta}\bigr](x';x') - i\bigl[{}^{\mu(\rho} 
\Delta^{\sigma)\nu}\bigr](x';x') \Bigr\} \; , \label{4ptFF} \qquad \\
\lefteqn{ -i \bigl[{}^{\mu\nu} \Sigma^{\rho\sigma}_{\text{3pt}} \bigr](x;x')
= -\tfrac{\kappa^2}{2} (aa')^{D-4} \Bigl\{ \tfrac1{16} \eta^{\mu\nu} 
\eta^{\rho\sigma} i\bigl[{}^{\alpha\beta} \Delta^{\gamma\delta}\bigr](x;x') 
i\bigl[{}_{\alpha\beta} \Delta_{\gamma\delta}\bigr](x;x') } \nonumber \\
& & \hspace{1cm} - \tfrac14 \eta^{\mu\nu} i\bigl[{}^{\alpha\beta} 
\Delta^{\gamma\rho}\bigr](x;x') i\bigl[{}_{\alpha\beta} 
\Delta_{\gamma}{}^{\sigma}\bigr](x;x') \nonumber\\
& & \hspace{-0.5cm} - \tfrac14 \eta^{\rho\sigma} i\bigl[{}^{\alpha\mu} 
\Delta^{\gamma\delta}\bigr](x;x') i\bigl[{}_{\alpha}{}^{\nu} 
\Delta_{\gamma\delta}\bigr](x;x') \!+\! i\bigl[{}^{\alpha\mu} 
\Delta^{\gamma(\rho}\bigr](x;x') i\bigl[{}_{\alpha}{}^{\nu} 
\Delta_{\gamma}{}^{\sigma)}\bigr](x;x') \Bigl\} . \quad \label{3ptFF}
\end{eqnarray}
We remind the reader that these results are not expressed in a manifestly 
covariant form (although they are covariant) because indices are raised 
and lowered with the Minkowski metric as opposed to the de Sitter background
metric. It should also be clear from expressions (\ref{3ptExpr}) and 
(\ref{Vvert}) that the last term in (\ref{3ptFF}) is only symmetrized with
respect to the indices $\rho$ and $\sigma$. 

Because the entire 4-point contribution (\ref{4ptFF}) is proportional 
to a delta function, we can simplify it using the coincidence limits,
\begin{align}
i\bigl[{}_{\alpha\beta} \Delta_{\gamma\delta}\bigr](x';x')  
& = -8H^2 a'^4 B'(0) \eta_{\alpha[\gamma}\eta_{\delta]\beta} \; , 
\label{coinFFcor} \\[10pt]
\left\{\partial_{\rho}i\bigl[{}_{\alpha\beta} \Delta_{\gamma\delta} \bigr]
\right\}(x';x') 
& = -16H^3a'^5B'(0) \nonumber \\
& \times \left({\delta^0}_{\rho}\eta_{\alpha[\gamma}\eta_{\delta]\beta}
+{\delta^0}_{[\alpha}\eta_{\beta][\gamma}\eta_{\delta]\rho}
+{\delta^0}_{[\delta}\eta_{\gamma][\beta}\eta_{\alpha]\rho}\right) \; . 
\label{derivCoin}
\end{align}
Here the function $B(a a' H^2 \Delta x^2)$ is the de Sitter invariant 
propagator of a minimally coupled scalar with mass $(D-2) H^2$,
\begin{equation}
B(y) = \tfrac{H^{D-2}}{(4\pi)^{\frac{D}2}} \Bigl\{ \tfrac{\Gamma(\frac{D}2)}{
\frac{D}2 - 1} (\tfrac{4}{y})^{\frac{D}2 -1} + \sum_{n=0}^{\infty} \Bigl[
\tfrac{\Gamma(n+\frac{D}2)}{\Gamma(n+2)} (\tfrac{y}{4})^{n-\frac{D}2 +2} -
\tfrac{\Gamma(n + D - 2)}{\Gamma(n + \frac{D}2)} (\tfrac{y}{4})^{n}\Bigr] 
\Bigr\} \; . \label{Bdef}
\end{equation}
In dimensional regularization the coincidence limit of its derivative 
is finite,
\begin{equation}
B'(0) = -\tfrac{H^{D-2}}{4 (4\pi)^{\frac{D}2}} \tfrac{\Gamma(D-1)}{
\Gamma(\frac{D}2 + 1)} \; . \label{Bprime}
\end{equation}
The coincidence limit (\ref{coinFFcor}) allows us to express the 4-point 
contribution as,
\begin{eqnarray}
\lefteqn{
-i\bigl[{}^{\mu\nu}\Sigma^{\rho\sigma}_{\text{4pt}}\bigr](x;x') = 
-\kappa^2 a^D H^2 B'(0) \Bigl\{ 
\tfrac12 \left( D^2 \!-\! 9D \!+\! 12 \right) \eta^{\mu(\rho}
\eta^{\sigma)\nu} } \nonumber\\
& & \hspace{4cm}
- \tfrac14 \left( D^2 \!-\! 9D \!+\! 16 \right) \eta^{\mu\nu}
\eta^{\rho\sigma} \Bigr\} i\delta^D(x-x') \; . \label{4ptlocal}
\end{eqnarray}
This latter form of the 4-point contribution will yield the cleanest 
computation of its conservation in the next section.

\section{Conservation}

The purpose of this section is to demonstrate that the sum of the two
primitive contributions (\ref{3ptFF}) and (\ref{4ptlocal}) is conserved,
even at $x^{\mu} = {x'}^{\mu}$. We begin by defining the ``Ward operator''
which represents the $h_{\mu\nu} = 0$ part of the divergence operator acting
on the variation of the matter action with respect to the graviton field.
This operator is then acted on the two primitive contributions and the sum 
is shown to be finite and vanishing in $D=4$ dimensions. The section closes 
by demonstrating that the stress tensor induced by integrating out the photon
field has exactly the same property.

\subsection{No Obstacle}

In order to state conservation, the end result of the Ward operator acting 
on the graviton self-energy must vanish in $D=4$. Here, we define the Ward 
operator as,
\begin{equation}\label{WardDeinition}
    {\mathcal{W}^\nu}_{\alpha\beta} \equiv {\delta^\nu}_{(\alpha}\partial_{\beta)} 
    + aH{\delta^\nu}_0 \eta_{\alpha\beta} \; .
\end{equation}
Acting the Ward operator on the 4-point contribution from (\ref{4ptlocal}) 
is simple,
\begin{align}
    {\mathcal{W}^\nu}_{\alpha\beta} &\times -i\bigl[{}^{\alpha\beta}
    \Sigma^{\rho\sigma}_{\text{4pt}}\bigr](x;x') \nonumber \\
    &= \tfrac{\kappa^2}{4}H^3a'^{D+1}B'(0) (D\!-\!1) (D\!-\!4) (D\!-\!6) 
    {\delta^\nu}_0 \eta^{\rho\sigma} i \delta^D(x-x') \nonumber \\
    &-\kappa^2 H^2 a'^D B'(0) \Bigl\{\tfrac{1}{2}(D^2 \!-\!9D \!+\! 12) 
    \eta^{\nu(\rho} \partial^{\sigma)} \!-\! \tfrac{1}{4} (D^2 \!-\!9D \!+\! 16)
    \eta^{\rho\sigma} \partial^{\nu} \Bigr\}  \nonumber \\
    &\times i\delta^D(x-x') \; . \label{4ptWard}
\end{align}
Meanwhile, the only terms which survive acting the Ward operator on the 
3-point contribution stem from the delta-function contribution of the field 
equations of the field-strength correlator, 
\begin{eqnarray}
\lefteqn{{\mathcal{W}^\nu}_{\alpha\beta} \!\times\! -i\bigl[{}^{\alpha\beta}
\Sigma^{\rho\sigma}_{\text{3pt}}\bigr](x;x') } \nonumber \\
& & \hspace{1cm} = \tfrac{\kappa^2}{2}(aa')^{D-4} \biggl[ \left\{ \partial_\beta 
i \bigl[{}^{\beta\alpha} \Delta^{\gamma(\rho}\bigr] \right\}_{\text{local}}(x;x')
\!\times\! i \bigl[{_{\alpha}}^{\nu} {\Delta_{\gamma}}^{\sigma)}\bigr](x;x') 
\nonumber \\
& & \hspace{1cm} -  \tfrac{1}{4}\eta^{\rho\sigma} \left\{ \partial_\beta 
i \bigl[{}^{\beta\alpha} \Delta^{\gamma\delta}\bigr] \right\}_{\text{local}}(x;x')
\!\times\! i \bigl[{_{\alpha}}^{\nu} \Delta_{\gamma\delta}\bigr](x;x') \biggr] 
\; . \qquad
\end{eqnarray}

The explicit form of the delta function contribution is, 
\begin{equation}
    \left\{ \partial_\mu i \bigl[{}^{\mu\nu} \Delta^{\rho\sigma}\bigr] 
    \right\}_{\text{local}}(x;x') = \frac{1}{a^{D-4}}\left( \eta^{\nu\sigma}
    \partial^{'\rho}- \eta^{\nu\rho}\partial^{'\sigma} \right) i \delta^D(x-x') \; .
\end{equation}
The most useful form of the result is obtained by reflecting the derivatives, 
and then by partially integrating each term. This allows us to apply the 
coincident identities of the propagators (\ref{coinFFcor}) and (\ref{derivCoin}). 
Then, the end-result of the Ward operator acting on the 3-point contribution is,
\begin{align}
    {\mathcal{W}^\nu}_{\alpha\beta}&\times -i\bigl[{}^{\alpha\beta}
    \Sigma^{\rho\sigma}_{\text{3pt}}\bigr](x;x') \nonumber \\
    &= \kappa^2H^3a'^{D+1}B'(0)(D-1)(D-4){\delta^\nu}_0\eta^{\rho\sigma}i\delta^D(x-x') \nonumber \\
    &-\kappa^2H^2a'^DB'(0)\left\{2(D-2)\eta^{\nu(\rho}\partial^{\sigma)}-(D-3)
    \eta^{\rho\sigma}\partial^{\nu}\right\}  \nonumber \\
    &\times i\delta^D(x-x') \; . \label{3ptWard}
\end{align}

Combining (\ref{3ptWard}) with (\ref{4ptWard}) yields the total obstacle to conservation,
\begin{align}
    {\mathcal{W}^\nu}_{\alpha\beta}&\times -i\bigl[{}^{\alpha\beta}
    \Sigma^{\rho\sigma}_{\text{tot}}\bigr](x;x') \nonumber \\
    &= \frac{\kappa^2}{4}(D-4)(D-2)(D-1)H^3a'^{D+1}B'(0){\delta^\nu}_0
    \eta^{\rho\sigma}i\delta^D(x-x') \nonumber \\
    &+\kappa^2(D-4)(D-1)H^2a'^DB'(0)\left\{\frac{1}{4}\eta^{\rho\sigma}
    \partial^{\nu}-\frac{1}{2}\eta^{\nu(\rho}\partial^{\sigma)}\right\} 
    i\delta^D(x-x') \; . \label{obstacle} 
\end{align}
Expression (\ref{Bprime}) gives $B'(0)$, which is finite in $D=4$ dimensions. Because
the obstacle (\ref{obstacle}) is proportional to $D-4$, it vanishes in $D=4$. Hence 
there is no obstacle to conservation, and thus there is no need to perform finite 
renormalization of the cosmological constant with a counterterm. However, it is worth 
mentioning that the result can be made null for any dimension $D$ using the finite
counterterm, 
\begin{equation}
    \Delta\mathcal{L} = -(D\!-\!4) (D\!-\!1) H^2 B'(0) \sqrt{-g} \; .
\end{equation}

\subsection{No Induced Stress Tensor}

The lack of an obstacle (\ref{obstacle}) to conservation in $D=4$ dimensions 
is related to the vanishing of the stress tensor induced by integrating out 
the photon field. The stress tensor operator is found by functionally 
differentiating the electromagnetic action with respect to the metric, 
\begin{equation}
    T_{\mu\nu} \equiv \tfrac{-2}{\sqrt{-g}} \tfrac{\delta S_{\text{EM}}}{\delta g^{\mu\nu}} 
    = F_{\mu\alpha} F_{\nu\beta} g^{\alpha\beta} -\tfrac{1}{4} g_{\mu\nu} g^{\rho\sigma} 
    g^{\alpha\beta} F_{\rho\alpha} F_{\sigma\beta} \; .
\end{equation}
Integrating out the differentiated photon fields amounts to replacing the
product of two field strengths by the coincidence limit of the field strength
correlator (\ref{coinFFcor}),
\begin{equation}
F_{\mu\alpha}(x) F_{\nu\beta}(x) \longrightarrow i[\mbox{}_{\mu\alpha} 
\Delta_{\nu\beta}](x;x) = -8 H^2 B'(0) g_{\mu [\nu} g_{\beta] \alpha} \; .
\end{equation}
This makes it simple to find,
\begin{equation}
T_{\mu\nu} \longrightarrow (D\!-\!4) (D\!-\!1) H^2 B'(0) g_{\mu\nu} \; .
\end{equation}
This stress tensor vanishes in $D = 4$ because expression (\ref{Bprime}) shows 
that $B'(0)$ is ultraviolet finite. Note the contrast with the minimally 
coupled scalar case, which shows a nonzero obstacle to conservation 
\cite{Tsamis:2023fri}, and a corresponding nonzero induced stress tensor 
\cite{Miao:2024nsz}.

\section{Gravitational Radiation}

Because there is no obstacle to conservation, the previous result for the
graviton self-energy is correct \cite{Wang:2015eaa}. In this section we
use the Schwinger-Keldysh \cite{Chou:1984es,Jordan:1986ug,Calzetta:1986ey,
Ford:2004wc} version of this result to quantum-correct the linearized 
Einstein equation,
\begin{equation}
\mathcal{D}^{\mu\nu\rho\sigma}\kappa h_{\rho\sigma}(x) = 8\pi G T^{\mu\nu}(x) 
+ \int \!\! d^4x' \, [\mbox{}^{\mu\nu} \Sigma^{\rho\sigma}](x;x')\!\times\!
\kappa h_{\rho\sigma}(x') \; . \label{linEin}
\end{equation}
The section begins by explaining the various symbols in this equation. We
then set the stress tensor to zero and solve for the leading late time 
1-loop correction to gravitational radiation.

On de Sitter background the Lichnerowicz operator is,
\begin{eqnarray}
\lefteqn{\mathcal{D}^{\mu\nu\rho\sigma} h_{\rho\sigma} = \tfrac12 a^2 
\Bigl[\partial^2 h^{\mu\nu} \!-\! \eta^{\mu\nu} \partial^2 h \!+\! 
\eta^{\mu\nu} \partial^{\rho} \partial^{\sigma} h_{\rho\sigma} \!+\! 
\partial^{\mu} \partial^{\nu} h \!-\! 2 \partial^{\rho} 
\partial^{(\mu} h^{\nu)}_{~~\rho} \Bigr] } \nonumber \\
& & \hspace{1cm} + a^3 H \Bigl[ \eta^{\mu\nu} \partial_0 h \!-\! \partial_0
h^{\mu\nu} \!-\! 2 \eta^{\mu\nu} \partial^{\rho} h_{\rho 0} \!+\! 2 
\partial^{(\mu} h^{\nu)}_{~~0} \Bigr] \!+\! 3 a^4 H^2 \eta^{\mu\nu} h_{00} 
\; , \qquad \label{Lichnerowicz}
\end{eqnarray}
where $h^{\mu}_{~\nu} \equiv \eta^{\mu\rho} h_{\rho\nu}$ and $h \equiv 
\eta^{\rho\sigma} h_{\rho\sigma}$. The Schwinger-Keldysh version of the
graviton self-energy is,
\begin{equation}
[\mbox{}^{\mu\nu} \Sigma^{\rho\sigma}](x;x') = -\tfrac{\kappa^2}{2^8 \cdot
5 \cdot \pi^3} \, \mathcal{C}^{\alpha\beta\gamma\delta\mu\nu} \!\times\!
{\mathcal{C}'}_{\alpha\beta\gamma\delta}^{~~~~~\rho\sigma} \Bigl[ 8 \pi 
\ln(a a') \delta^4(x \!-\! x') \!+\! f_B(x \!-\! x')\Bigr] . \label{retSKSE}
\end{equation}
Here the 2nd order tensor differential operator $\mathcal{C}_{\alpha\beta
\gamma\delta}^{~~~~~ \mu\nu}$ is obtained by expanding the Weyl tensor of the
conformally transformed metric $\widetilde{g}_{\mu\nu} \equiv \eta_{\mu\nu}
+ \kappa h_{\mu\nu}$,
\begin{equation}
\widetilde{C}_{\alpha\beta\gamma\delta} \equiv \mathcal{C}_{\alpha\beta
\gamma\delta}^{~~~~~\mu\nu} \!\times\! \kappa h_{\mu\nu} + O(\kappa^2 h^2)
\; . \label{Cdef}
\end{equation}
It is manifestly traceless and transverse, and its explicit form can be 
found in \cite{Park:2011ww,Leonard:2014zua}. The function $f_B(x - x')$
is \cite{Miao:2024pwd},
\begin{align}
    f_B(x \!-\! x') \equiv \partial^4 \Bigl\{ \theta(\Delta \eta \!-\! \Delta r)
    \Bigl(\ln[\mu^2 (\Delta \eta^2 \!-\! \Delta r^2)] \!-\! 1 \Bigr) \Bigr\} 
    \; , \label{fB}
\end{align}
where $\Delta \eta \equiv \eta - \eta'$ and $\Delta r \equiv \Vert \vec{x} -
\vec{x}'\Vert$. Note that the Schwinger-Keldysh graviton self-energy is both 
real and causal, albeit nonlocal.

We want to set the stress tensor to zero in equation (\ref{linEin}) and solve 
for 1-loop corrections to gravitational radiation. A plane gravitational wave 
can be characterized by its wave vector $\vec{k}$ and a polarization tensor 
$\epsilon_{\mu\nu}$ which is purely spatial ($\epsilon_{00} = 0 = \epsilon_{0i}$), 
transverse ($k_i \epsilon_{i j} = 0$) and traceless ($\epsilon_{ii} = 0$),
\begin{equation}
\kappa h_{\mu\nu}(\eta,\vec{x}) = \epsilon_{\mu\nu} e^{i \vec{k} \cdot \vec{x}}
u(\eta,k) \; . \label{gengrav}
\end{equation}
The Lichnerowicz operator acts on (\ref{gengrav}) to give,
\begin{equation}
\mathcal{D}^{\mu\nu\rho\sigma} \kappa h_{\rho\sigma} = \epsilon^{\mu\nu} 
e^{i\vec{k} \cdot \vec{x}} \!\times\! -\tfrac{a^2}{2} [\partial_0^2 \!+\! 
2 a H \partial_0 \!+\! k^2] u(\eta,k) \; . \label{LHS}
\end{equation}
One can also obtain a general result for any self-energy of the form
(\ref{retSKSE}) \cite{Leonard:2014zua},
\begin{eqnarray}
\lefteqn{ \int \!\! d^4x' \, \mathcal{C}^{\alpha\beta\gamma\delta\mu\nu}
\!\times\! {\mathcal{C}'}_{\alpha\beta\gamma\delta}^{~~~~~\rho\sigma} 
F(\eta,\eta',\vec{x} \!-\! \vec{x}') \!\times\! \kappa h_{\rho\sigma}(x') }
\nonumber \\
& & \hspace{1cm} = 2 \partial_{\rho} \partial_{\sigma} \!\! \int \!\! d^4x' \, 
F(\eta,\eta',\Delta \vec{x}) \!\times\! \widetilde{C}_{\rm lin}^{\rho\mu
\sigma\nu}(x') \; , \qquad \\
& & \hspace{1cm} = \epsilon^{\mu\nu} e^{i \vec{k} \cdot \vec{x}} \Bigl\{ 
-\tfrac12 (\partial_0^2 \!-\! k^2) \!\! \int \!\! d^4x' \, F(\eta,\eta',
\Delta \vec{x}) (\partial_0^{\prime 2} \!-\! k^2) u(\eta',k) 
e^{-i \vec{k} \cdot \Delta \vec{x}} \nonumber \\
& & \hspace{4cm} -2 k^2 \partial_0 \!\! \int \!\! d^4x' \, F(\eta,\eta',
\Delta \vec{x}) \partial_0' u(\eta',k) e^{-i\vec{k} \cdot \Delta \vec{x}}
\Bigr\} . \qquad \label{RHS}
\end{eqnarray}
Equating (\ref{LHS}) to (\ref{RHS}) gives the graviton mode equation,
\begin{eqnarray}
\lefteqn{ a^2 [\partial_0^2 \!+\! 2 a H \partial_0 \!+\! k^2] u(\eta,k) }
\nonumber \\
& & \hspace{1cm} = (\partial_0^2 \!-\! k^2) \!\! \int \!\! d^4x' \, 
F(\eta,\eta',\Delta \vec{x}) (\partial_0^{\prime 2} \!-\! k^2) u(\eta',k) 
e^{-i \vec{k} \cdot \Delta \vec{x}} \nonumber \\
& & \hspace{4.5cm} + 4 k^2 \partial_0 \!\! \int \!\! d^4x' \, F(\eta,\eta',
\Delta \vec{x}) \partial_0' u(\eta',k) e^{-i\vec{k} \cdot \Delta \vec{x}}
\; . \qquad \label{modeEQN}
\end{eqnarray}

Because we only know the graviton self-energy at 1-loop order the
only way to solve equation (\ref{modeEQN}) is perturbatively,
\begin{equation}
u(\eta,k) \equiv u_0(\eta,k) + \kappa^2 u_1(\eta,k) + \kappa^4 u_2(\eta,k)
+ \dots \; . \label{uexp}
\end{equation}
The 0th order solution is,
\begin{equation}
u_0(\eta,k) = \tfrac{H}{\sqrt{2 k^3}} [1 \!-\! \tfrac{i k}{aH} ] 
\exp[\tfrac{ik}{aH}] = \tfrac{H}{\sqrt{2 k^3}} [1 \!+\! i k \eta]
e^{-i k \eta} \; . \label{uzero}
\end{equation}
The derivatives inside the integrals of equation (\ref{modeEQN}) obey,
\begin{equation}
\partial_0 u_0 = \tfrac{H}{\sqrt{2 k^3}} \, k^2 \eta e^{-i k \eta} =
\tfrac{i}{2 k} (\partial_0^2 \!-\! k^2) u_0 \; . \label{duIDs}
\end{equation}
This means we can express the order $\kappa^2$ part of the mode equation 
(\ref{modeEQN}) as,
\begin{eqnarray}
\lefteqn{a^2 [\partial_0^2 \!+\! 2 a H \partial_0 \!+\! k^2] u_1(\eta,k) }
\nonumber \\
& & \hspace{0cm} = \tfrac{i k (\partial_0 + i k)^2}{2^7 \cdot 5 \cdot \pi^3} 
\! \int \!\! d^4x' \Bigl[8 \pi \ln(a a') \delta^4(x \!-\! x') \!+\!
f_B(x \!-\! x')\Bigr] \partial'_0 u_0(\eta',k) e^{-i \vec{k} \cdot \Delta 
\vec{x}} \; . \qquad \label{u1EQN}
\end{eqnarray}

In addition to (\ref{duIDs}) the 0th order mode function obeys,
\begin{equation}
(\partial_0 \!+\! i k) \partial_0 u_0(\eta,k) = \tfrac{H}{\sqrt{2 k^3}} \, 
k^2 e^{-i k \eta} \qquad , \qquad (\partial_0 \!+\! i k)^2 \partial_0 
u_0(\eta,k) = 0 \; . \label{dduIDs}
\end{equation}
Of course we can perform the delta function integral in (\ref{u1EQN}),
and the nonlocal contribution vanishes when we reflect the derivatives
onto $\partial_0' u_0(\eta',k)$ and invoke the second identity of 
(\ref{dduIDs}),
\begin{equation}
a^2 [\partial_0^2 \!+\! 2 a H \partial_0 \!+\! k^2] u_1(\eta,k) = -
\tfrac{i k (\partial_0 + i k)^2}{2^4 \cdot 5 \cdot \pi^2} \Bigl[2 \ln(a)
\partial_0 u_0(\eta,k) \Bigr] \; . \label{step1}
\end{equation}
Acting the derivatives on the right hand side of (\ref{step1}) gives,
\begin{equation}
a^2 [\partial_0^2 \!+\! 2 a H \partial_0 \!+\! k^2] u_1(\eta,k) = -
\tfrac{i a H k^3}{2^3 \cdot 5 \cdot \pi^2} \times \tfrac{H}{\sqrt{2 k^3}}
e^{-i k \eta} \; . \label{step2} 
\end{equation}
The late time solution is easily recognized,
\begin{equation}
u_1(\eta,k) \longrightarrow \tfrac{H}{\sqrt{2 k^3}} \times 
\tfrac{i H^2 \ln(a)}{120 \pi^2} (\tfrac{k}{a H})^3 \; . \label{u1late}
\end{equation}

Using (\ref{u1late}) it is straightforward to see that the late time 
limiting form of the electric part of the Weyl tensor is,
\begin{equation}
C_{0i0j}(x) \longrightarrow C_{0i0j}^{\rm tree}(x) \Bigl\{1 \!+\!
\tfrac{\kappa^2 H^2}{40 \pi^2} \, \ln{(a)} + \dots \Bigr\} \; . 
\label{EMWeyl}
\end{equation}
It is worth recalling that the 1-loop correction to the Newtonian 
potential is \cite{Wang:2015eaa},
\begin{equation}
\Psi(t,r) = \tfrac{G M}{a r} \Bigl\{1 + \tfrac{\kappa^2}{60 \pi^2 a^2 r^2}
+ \tfrac{\kappa^2 H^2}{40 \pi^2} \ln(a H r) + \dots \Bigr\} \; .
\label{EMNewton}
\end{equation}
Like (\ref{Coulomb}), the fractional correction proportional to 
$\kappa^2/a^2 r^2$ is the de Sitter version of an effect well known on
flat space background \cite{Radkowski:1970,Capper:1974ed}. Interestingly, 
the fractional corrections proportional to $\kappa^2 H^2$ are the same in
expressions (\ref{EMWeyl}) and (\ref{EMNewton}). The positive sign is
opposite to the analogous effects (\ref{MMCrad}-\ref{MMCNewton})
engendered by a loop of massless, minimally coupled scalars 
\cite{Miao:2024atw}. One interpretation for the different sign is that 
energy must be drawn from the gravitational sector to produce 
inflationary scalars whereas inflation produces no photons and the 
redshift of their 0-point energies actually contributes energy to the
gravitational sector.

\section{Renormalization Group Resummation}

Gravity plus electromagnetism is not perturbatively renormalizable, even at
1-loop order \cite{Deser:1974zzd,Deser:1974cz}. However, the ultraviolet 
divergences of any quantum field theory can be subtracted off using BPHZ
counterterms (the initials stand for Bogoliubov, Parasiuk 
\cite{Bogoliubov:1957gp}, Hepp \cite{Hepp:1966eg} and Zimmermann 
\cite{Zimmermann:1968mu,Zimmermann:1969jj}). For matter plus gravity at
1-loop order these counterterms are \cite{tHooft:1974toh},
\begin{equation}
\Delta \mathcal{L} = c_1 R^2 \sqrt{-g} + c_2 C^{\alpha\beta\gamma\delta}
C_{\alpha\beta\gamma\delta} \sqrt{-g} \; . \label{DeltaL}
\end{equation}
In dimensional regularization the coefficients for Maxwell plus Einstein
are \cite{Wang:2015eaa},
\begin{equation}
c_1 = 0 \qquad , \qquad c_2 = \tfrac{\mu^{D-4}}{D - 4} \!\times\! 
\tfrac{\Gamma(\frac{D}2)}{2^8 \pi^{\frac{D}2}} \tfrac{(D-2) D}{(D+1) (D-3)^2}
\; . \label{c1c2}
\end{equation}

Gravitational effects such as (\ref{MMCrad}-\ref{MMCNewton}) which are 
induced by a loop of massless, minimally coupled scalars can be explained
by a variant of the renormalization group \cite{Miao:2024nsz}. Because the
counterterms have the same form (\ref{DeltaL}) as for Maxwell plus
Einstein, the analysis is identical. For general coefficients $c_1$ and
$c_2$ one finds that a certain combination can be viewed as a field strength
renormalization \cite{Miao:2024nsz},
\begin{equation}
\delta Z = D \Bigl[ 2 (D\!-\!1) c_1 - c_2\Bigr] \kappa^2 H^2 \; . \label{dZ}
\end{equation}
Substituting the values (\ref{c1c2}) relevant to Maxwell plus Einstein gives
the gamma function,
\begin{equation}
\gamma \equiv \tfrac{\partial \ln(1 + \delta Z)}{\partial \ln(\mu^2)} =
-\tfrac{\kappa^2 H^2}{80 \pi^2} \; . \label{gamma}
\end{equation}

The Weyl tensor (\ref{EMWeyl}) and the Newtonian potential (\ref{EMNewton})
are both 2-point Green's functions, for which the Callan-Symanzik equation 
implies,
\begin{equation}
\Bigl[ \tfrac{\partial}{\partial \ln(\mu)} + \beta_G \tfrac{\partial}{\partial G}
+ 2 \gamma \Bigr] G^{(2)} = 0 \; . \label{CSeqn}
\end{equation}
Because the photon propagator has no tail term, the only secular logarithms
from photon loops derive from renormalization (\ref{RGsource}). This means 
that factors of $\ln(\mu)$ are always come in the form $\ln(\mu a)$, so we can 
replace the derivative with respect to $\ln(\mu)$ by a derivative with respect 
to $\ln(a)$ in equation (\ref{CSeqn}). Taking account of the fact that the beta 
function vanishes at 1-loop gives a simple RG explanation for the laboriously 
derived results (\ref{EMWeyl}-\ref{EMNewton}).

\section{Conclusions}

We have re-examined electromagnetic corrections to gravity on de Sitter 
background, which were the subject of a previous study \cite{Wang:2015eaa}.
We were particularly concerned that there might be a local obstacle to
conservation of the graviton self-energy, as was recently found for massless,
minimally coupled scalar loop corrections \cite{Tsamis:2023fri}, and which
resulted in a small mistake in scalar corrections to the Newtonian potential
\cite{Leonard:2014zua,Park:2015kua}. In section 3 we demonstrate that there
is no such obstacle, which means that the previous result (\ref{EMNewton}) 
for electromagnetic corrections to the Newtonian potential is correct. In
section 4 we worked out the 1-loop corrections to the electric components 
of the Weyl tensor (\ref{EMWeyl}). 

It is worth commenting on the positive sign of the order $G H^2$ corrections 
to (\ref{EMWeyl}) and (\ref{EMNewton}). Because electromagnetism is conformally 
invariant in $D = 4$ dimensions, there is no production of photons during 
inflation, so the positive sign means that electromagnetic 0-point energy is 
strengthening gravity. This might be viewed as a flow of excitation from the
redshifting 0-point energy. In sharp contrast, the order $G H^2$ corrections
to (\ref{MMCrad}) and (\ref{MMCNewton}) induced by massless, minimally coupled
scalars are negative. A possible physical interpretation is that inflation does 
produce massless, minimally coupled scalars, and the energy for this must be
drawn from gravity.

In section 5 we showed that both of the electromagnetic corrections 
(\ref{EMWeyl}) and (\ref{EMNewton}) can be explained by a variant of the 
renormalization group. The same analysis can be used to resum these large 
logarithms to all orders, which we present here with the factors of $\hbar$ 
and $c$ restored,
\begin{eqnarray}
C_{0i0j} &\!\!\! = \!\!\!& C_{0i0j}^{\rm tree} \Bigl\{1 \!+\! 
\tfrac{2 \hbar G H^2}{5 \pi c^5} \, \ln[a] \!+\! \dots\Bigr\} \longrightarrow
C_{0i0j}^{\rm tree} \!\times\! [a(t)]^{\frac{2 \hbar G H^2}{5 \pi c^5}} , 
\qquad \label{Weyl} \\
\Psi &\!\!\! = \!\!\!& \tfrac{G M}{a r} \Bigl\{1 + 
\tfrac{4 \hbar G}{15 \pi c^3 a^2 r^2} \!+\! \tfrac{2 \hbar G H^2}{5 \pi c^5} 
\, \ln[\tfrac{a H r}{c}] \!+\! \dots \Bigr\} \longrightarrow \tfrac{G M}{a r}
\!\times\! [\tfrac{a(t) H r}{c}]^{\frac{2 \hbar G H^2}{5 \pi c^5}} .
\qquad \label{Newton}
\end{eqnarray}
Note from the fractional power of $r$ in the Newtonian potential that the
resulting late time effective field theory cannot be local. Note also that
``late time'' here means during an endless phase of de Sitter expansion. 
More work needs to be done to infer what would happen in the current 
universe after a long phase of primordial expansion, although a very good 
approximate solution has been obtained for nonlinear sigma models 
\cite{Kasdagli:2023nzj} which proves that significant late time effects 
can persist \cite{Woodard:2023cqi}.

Our renormalization group explanation adds to the growing list of 
large inflationary loop corrections induced by gravity on matter 
\cite{Glavan:2021adm,Glavan:2023tet}, and by matter on gravity 
\cite{Miao:2024atw}. This completes the analysis of Klein Gordon plus
Einstein and Maxwell plus Einstein. It remains to study Dirac plus
Einstein, and pure gravity. Graviton loop correction to fermions were 
treated in \cite{Miao:2006gj}, but have so far not been explained using 
the renormalization group, and fermion corrections to gravity have not 
been studied. Graviton loop corrections have been computed for 
gravitational radiation \cite{Tan:2021lza} and for the Newtonian 
potential \cite{Tan:2022xpn}. The preliminary steps have been taken 
to generalize Starobinsky's stochastic formalism to pure gravity
\cite{Miao:2024shs} but more work needs to be done.

With the possible exception of pure gravity, it seems as if all large 
loop corrections involving gravity plus matter arise from renormalization 
(\ref{RGsource}), with no contribution from the tail part of the 
propagator (\ref{tailterm}). This is no doubt related to the fact that
matter fields either possess no tail term, or have it suppressed by 
derivatives. However, there is an important distinction between photons,
which have no tail at all, and the massless, minimally coupled scalar, 
whose differentiated tail induces an obstacle to conservation of the 
graviton self-energy \cite{Tsamis:2023fri}. We therefore expect that
massless fermions, which have no tail, will induce no obstacle and that
their large loop corrections can be resummed using the renormalization
group. We further expect that there will be an obstacle for pure quantum
gravity (whose $\kappa h \partial h \partial h$ interaction involves an
undifferentiated field with a tail) and that Starobinsky's stochastic 
formalism will be necessary to explain its large loop corrections.

Because all large loop corrections in gravity plus matter seem to derive
from renormalization (\ref{RGsource}), it would be interesting to see if 
they persist for the contributions from massive particles. Of course 
renormalization will certainly induce such logarithms, but it is not yet
clear whether or not they will be suppressed by inverse powers of the 
scale factor. Such suppressed logarithms occur as well in other models
\cite{Miao:2021gic,Glavan:2021adm,Glavan:2023tet,Miao:2024atw}.

\vskip .5cm

\centerline{\bf Acknowledgements}

This work was partially supported by NSF grant PHY-2207514 
and by the Institute for Fundamental Theory at the University of Florida.

\end{document}